\documentclass{iau} 
\usepackage{graphicx}

\title[Low-Z massive stars] 
{Low-metallicity (sub-SMC) massive stars}

\author[M. Garcia, A. Herrero, \etal]
{Miriam Garcia$^1$, Artemio Herrero$^{2,3}$, \\
  \and \\
  Francisco Najarro$^1$, In\'es Camacho$^{2,3}$, Daniel J. Lennon$^4$, Miguel A. Urbaneja$^5$, Norberto Castro$^6$}

\affiliation{$^1$Centro de Astrobiolog\'{\i}a (INTA-CSIC), Departamento de Astrof\'{\i}sica. \\
Ctra. Torrej\'on a Ajalvir km.4, E-28850 Torrej\'on de Ardoz (Madrid), Spain \\
email: {\tt mgg@cab.inta-csic.es} \\[\affilskip]
$^2$Instituto de Astrof\'{\i}sica de Canarias, 38205 La Laguna (Tenerife), Spain\\
$^3$Departamento de Astrof\'{\i}sica, Universidad de La Laguna, 38206 La Laguna (Tenerife), Spain\\
$^4$European Space Astronomy Centre (ESA/ESAC), Villanueva de la Ca\~nada (Madrid), Spain\\
$^5$Institut fuer Astro- und Teilchenphysik, Universitaet Innsbruck, Innsbruck, Austria\\
$^6$Astronomy Department, University of Michigan, Ann Arbor, MI 48109, USA
}

\pubyear{2016}
\volume{329}  
\jname{The lives and death-throes of massive stars}
\editors{A.C. Editor, B.D. Editor \& C.E. Editor, eds.}

\usepackage{graphicx}
\usepackage[]{natbib}
\usepackage{txfonts}
\newcommand{\Msun}{$\rm M_{\odot}$}
\newcommand{\Osun}{$\rm O_{\odot}$}
\newcommand{\Fesun}{$\rm Fe_{\odot}$}
\newcommand{\Zsun}{$\rm Z_{\odot}$}

\newcommand{\Mini}{\mbox{$M_{ini}$}}

\newcommand{\Teff}{\mbox{$T_{\rm eff}$}}
\newcommand{\teff}{\mbox{$T_{\rm eff}$}}

\newcommand{\vinf}{\mbox{$v_{\infty}$}}
\newcommand{\vesc}{\mbox{$v_{esc}$}}

\newcommand{\Mdot}{$\dot M$}

\newcommand{\vrot}{$v_{rot}$}

\newcommand{\vini}{$v_{ini}$}

\newcommand{\kms}{~$\rm km \, s^{-1}$}
\newcommand{\myr}{M$_{\odot}$ yr$^{-1}$}

\newcommand{\halpha}{$\rm H_{\alpha}$}

\begin{document}

\maketitle

\begin{abstract}
  The double distance and metallicity frontier marked by the SMC has been finally broken with the aid of
  powerful multi-object spectrographs installed at 8-10m class telescopes.
  VLT, GTC and Keck have enabled studies of massive stars in dwarf irregular galaxies
  of the Local Group with poorer metal-content than the SMC.
  The community is working to test the predictions of evolutionary models in the low-metallicity regime,
  set the new standard for the metal-poor high-redshift Universe,
  and test the extrapolation of the physics of massive stars to environments
  of decreasing metallicity.
  In this paper, we review current knowledge on this topic.
\end{abstract}

\firstsection 
\section{Introduction}
\label{s:intro}

Massive stars leave their imprint through the ages of the Universe and, as such, hold the key
to interpret a plethora of astrophysical phenomena.
The copious amount of ionizing and mechanical feedback from a population of massive stars
drives locally the gas dynamics of their natal cloud,
and can impact at a global level the evolution of host galaxies.
Either as individuals or in binary systems,
massive stars are the progenitors of the most energetic events in the Universe
that can be used to probe high-redshifts: type Ibc,II supernovae (SNe)
and arguably pair-instability supernovae, super-luminous supernovae and long $\gamma$-ray bursts (GRBs).
The aftermath products, neutron stars and black holes, are sites of extreme physics.

In order to understand and quantify the role of massive stars in an evolving Universe,
and to eventually use SNe/GRBs as lighthouses,
it is necessary to describe the variation of their physical properties
as a function of metallicity.
The metal-poor regime is subject to a particularly growing interest in order to
understand the conditions of earlier cosmic epochs, and to ultimately
extrapolate the prescriptions for physical properties 
to the roughly metal-free Universe
at the re-ionization epoch.

We do expect significant differences between metal poor (Z$\rm \leq$1/10\Zsun) massive stars
and those in the Milky Way (MW) in terms that will translate into significant impact on their feedback.
Radiation-driven winds (RDW) are ubiquitous in the hot stages of massive star evolution.
They not only drive mechanical feedback but also regulate the evolution
of the star \citep[e.g.][]{Mal94}, and therefore
the overall ionizing radiation and the final properties of the pre-SN core.
Because the driving mechanism involves absorption and re-emission of photons by metallic transitions
RDWs are expected to decrease with decreasing metallicity, and be almost negligible
at Z$\sim$1/1000\Zsun~ except for very bright objects \citep{K02}.
However, the metallicity dependence of RDWs
has only been confirmed observationally down to the
Small Magellanic Cloud (SMC) \citep{Mal07b}.

The evolutionary pathways may also change drastically.
Metal-poor massive stars rotating sufficiently fast may bring
He produced at the stellar core to the surface,
mix it in fast time-scales, and experience chemical homogeneous evolution (CHE).
The star will never leave the high temperature regime of the HR-diagram (HRD);
the overall ionizing energy emitted through its evolution
will be enhanced with respect to analog stars
that, either because of a slower initial \vrot,
or a higher metallicity powered wind,
do reach the red supergiant stage (RSG).
For instance, a 150\Msun~ undergoing CHE will double the amount of
HI-ionizing photons and quadruple the HeII-ionizing photons
with respect to the redwards evolving analog
according to \citet{Szal15}'s 1/50\Zsun~ tracks.
CHE of binary massive stars within the Roche lobe
is one proposed scenario to generate the double
$\sim$30\Msun~ black hole system that produced the
first detection of gravitational waves when it merged \citep{MM16}.
Yet, while some evidence on CHE exists \citep{MDRM13}
a sufficiently large number of stars that establishes this
evolutionary pathway has not been detected.

Today, the Small Magellanic Cloud stands
the reference for the metal-poor Universe.
Spectral libraries of SMC stars feed population synthesis models \citep[e.g.][]{LLH01}.
However, the SMC's 1/5\Zsun~ falls short of the $\sim$1/30\Zsun~ metallicity 
at the peak of star formation \citep{MD14} and is
not valid an approximation for the roughly metal-free early Universe.
In this paper we will review recent efforts to surpass the SMC frontier,
and what we have learnt about sub-SMC metallicity massive stars.

\begin{table}[t]
  \begin{center}
  \caption{Metal-poor early-type massive stars: headcount}
  \label{T:census}
 {\scriptsize
  \begin{tabular}{|l|r r r r|}\hline 
{\bf       }   & {\bf IC~1613} & {\bf NGC~3109} & {\bf ~~WLM~~}   & {\bf Sextans~A} \\  \hline
Metallicity    & 1/7 \Osun     & 1/7 \Osun      & 1/7 \Osun       & 1/10 \Zsun      \\
\# O-B2 stars  & 56~~~~~~~     & 44~~~~~~~      & 8~~~~~~~        & 12~~~~~~~       \\
\# O-stars     & 26~~~~~~~     & 12~~~~~~~      & 2~~~~~~~        &  5~~~~~~~       \\
\# Earlier than O7 &  4~~~~~~~     &  0~~~~~~~      & 0~~~~~~~        &  0~~~~~~~       \\  \hline
  \end{tabular}
  }
 \end{center}
\vspace{1mm}
 \scriptsize{
   {\it References:}
   \citet{ALM88,Lal02,Bal06,Bal07,Eal07,GH13a,Cal16}, Herrero, Garcia \etal\ in prep.
  }
\end{table}

\section{The 1/7 \Osun~ galaxies IC~1613, NGC~3109 and WLM}
\label{s:ic1613}

\begin{figure}[t]
\begin{center}
 \includegraphics[width=\textwidth]{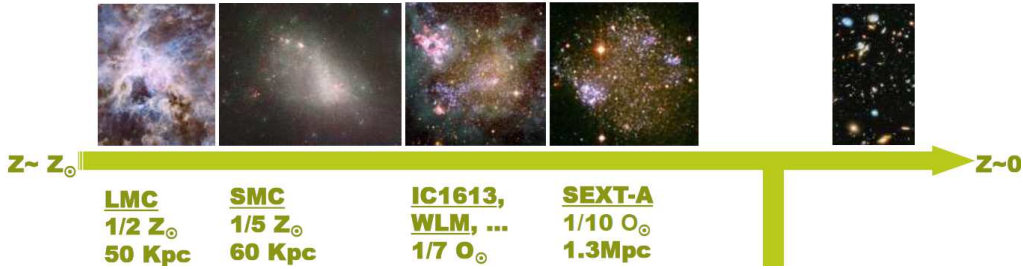} 
 \caption{Road-map to the early-Universe: sequence of resolved galaxies of decreasing metal content.
   Characterizing massive stars in this galaxies will open the metal-poor regime.
   The ultimate goal is to produce prescriptions valid for the
   early-Universe.
  }
   \label{F:road}
\end{center}
\end{figure}

The full development of the 8-10m telescopes and their multi-object spectrographs (MOS) in the
mid-2000's enabled the discovery and study of massive stars beyond the Magellanic Clouds with good quality data.
Three galaxies were identified within reach,
with promising 1/7\Osun\footnote{
At this point, we would like to note that metallicity (Z/\Zsun)
is often calculated from the ratio of oxygen abundances against the Solar value only,
with the implied hypothesis that the chemical mixture follows the Solar pattern.
However, the relative ratios of elements (e.g. $\alpha$/Fe) do vary with the chemical evolution of galaxies (as we discuss again in 
Sect.~\ref{ss:meta}).
For this reason, we use the notation of O/\Osun~ and Fe/\Fesun~ to distinguish the metallicity
indicator, and only use Z/\Zsun~ when information from at least both Fe and O is available.
  }
nebular abundances: IC~1613, NGC~3109 and WLM
(Fig.\,\ref{F:road}).

Spectroscopic studies with the previous generation of 4m telescopes
had been restricted to the brightest supergiants, LBV and WR stars,
because of faint magnitude limits and crowding issues.
Nowadays the 10.4m Gran Telescopio Canarias (GTC)
enables R$\sim$800 identification spectroscopy of an IC~1613 O-type star at 750Kpc (V=19.6) in 1 hour.
The actual exposure times per target are being decreased with detectors of ever increasing sensitivity,
and divided by the multiplexing capabilities of the MOSs.

The task first required a strong exploratory effort
to confirm the nature of massive star candidates with spectroscopy.
The candidates could not be directly targeted with multi-slit or multi-fiber spectrographs
for deep high resolution  data because blue massive stars cannot
be solely identified from optical colors \citep[though some optimized methods for the blue types exist, e.g.][]{GH13a}.
The high multi-plexing of VLT-VIMOS and FORS was of great help, although
numbers are still scarce.
Table \ref{T:census} summarizes the number of sub-SMC metallicity early-type massive stars known to date.

Although great progress has been made, our studies are still limited in sample size, resolution and epochs.
At the moment, we cannot tackle very interesting issues like rotational velocities, binary frequencies,
or the possibility of a top-heavy initial mass function in metal-poor environments
(in connection to the extremely massive, metal-free First Stars).
However, the observations do allow for quantitative analyses to obtain stellar
parameters, constrain the winds and enable first contrast against evolutionary
models. 

\subsection{\Teff~ \textit{vs} Spectral Type calibration}

\begin{figure}[t]
\begin{center}
 \includegraphics[width=\textwidth]{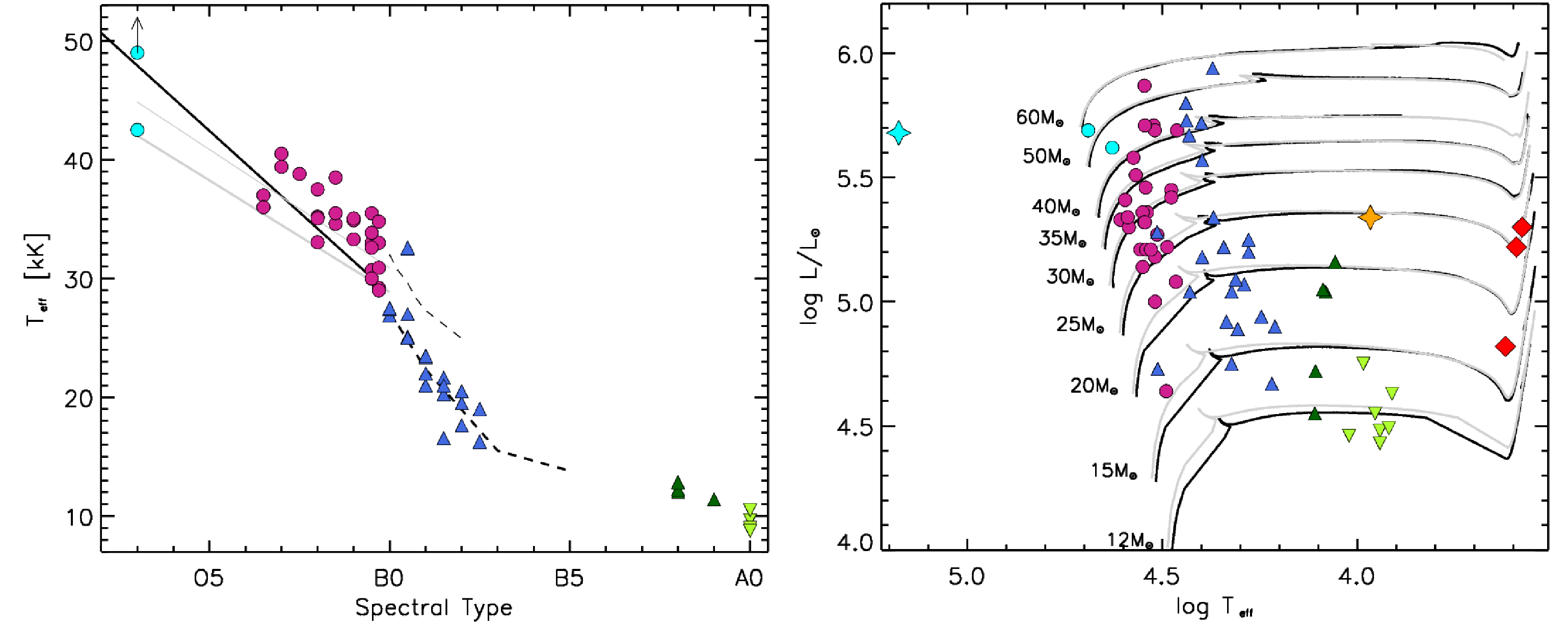} 
 \caption{Properties of known stars in 1/7\Osun~ galaxies.
   \textbf{Left:} \Teff~ \textit{vs} spectral type for early-type (O-A0) massive stars,
   compared with SMC results: \citet{Tral07}'s calibration for SMC B-stars (black dashed lines),
   and a linear fit to \citet{Mal07b} and \citet{PMal09} parameters for O stars (black solid line).
   For comparison, the MW \Teff~ \textit{vs} spectral type calibration by \citet{MSH05} is included in gray.
   For all calibrations: thick lines represent supergiants and thin lines dwarfs.
   \textbf{Right:} stars in the HR-diagram compared with \citet{Brott1}'s SMC tracks with \vini=300\kms (black) and 0\kms (gray).
   {\bf References:} Stellar samples collected from
   \citet[][in prep.]{Bal07,Eal07,Tal07,GH13a,Tal13,Tal14,Hal14,Bal15,Hal10,Hal12}, Camacho \etal\ in prep.
         {\bf Symbols:} Circles: O-stars, early-types in cyan; Up-triangles: B-supergiants, late-types in dark-green;
         Down-triangles: A-supergiants; Red-rhomboids: RSGs; Orange-star: the LBVc V39; Cyan-star: the WO DR1.
 }
   \label{F:HRD}
\end{center}
\end{figure}

The effective temperatures (\teff) of all OBA stars studied in IC~1613, NGC~3109 and WLM
are compiled in Fig.\,\ref{F:HRD}-left, and compared against calibrations with spectral type
for various metallicities.
These calibrations are useful to assign effective temperatures
in sight of spectral type to first approximation,
and to quantify ionizing photons emitted by unresolved populations.

The temperatures of O-stars show a large dispersion in Fig.\,\ref{F:HRD}-left.
The scatter is of the order of the expected differences between different luminosity
classes, although it likely reflects the heterogeneity of the analyzed data that either
have low resolution, low signal-to-noise ratio (SNR) or both.
When only giant and supergiant stars are considered, the 1/7\Osun~ effective temperature scale
is roughly $\sim$1000~K hotter than the SMC's \citep{GH13a}.

B-supergiants show a much smaller dispersion, and roughly follow
\citet{Tral07}'s calibration for the SMC. The early-A supergiants
can be found at the extrapolation of this calibration.
One early-B supergiant stands out with high
effective temperature. This is likely a miss-classification issue,
since low resolution or SNR may hinder the detection of the
HeII lines, which will be weak at the latest O-types.

\subsection{The HR-diagram}

The 1/7 \Osun~ massive stars are found at their expected loci in the HR-diagram (Fig.\,\ref{F:HRD}-right).
The types that have so far been confirmed by spectroscopy include O-stars and BA supergiants,
plus some advanced stages in IC~1613: RSGs, the oxygen Wolf-Rayett (WO) star DR1, and the LBV-candidate V39.
O-type stars draw the main sequence, with only one early-O star close to the zero-age main sequence (ZAMS).
B-supergiants overlap with the H-burning region of SMC tracks,
which could be partly explained by the shift of
the evolutionary paths to higher \teff~ as metallicity decreases.

The total sample is still small but at present there is no evidence
of a sequence of stars undergoing CHE, which would be located at higher temperatures
than the ZAMS. Likewise, no super-luminous RSG has been detected that would support
the inflation and redwards evolution scenario of slowly-rotating very metal-poor stars \citep{Szal15}.
At the moment we cannot contrast this information with rotational velocities
because of the low-resolution used for the majority of the observing runs,
but no high-speed rotator has been detected.

Finally, it is noteworthy that the most massive stars detected,
including the WO star in IC~1613 \citep{Tral15},
have \Mini$\sim$60\Msun.

\subsection{The Wind Momentum - Luminosity Relation (WLR)}
\label{ss:wlr}

Mass loss is one crucial ingredient to compute the evolution and fate of massive stars.
Quantifying true mass loss rates, providing prescriptions that hold
in varying metallicity environments, and testing such prescriptions in metal-poor environments,
is a must to be able to simulate massive stars at high-redshifts, their death and feedback.

The WLR is currently the main diagnostic tool, and compares observational
measurements of the mechanical wind-momentum ($\propto$ \Mdot $\cdot$ \vinf) with theory.
Based on the nature of RDWs,
the WLR tests simultaneously the expected positive correlation
of wind-momentum with stellar luminosity and its anti-correlation with metallicity.

\begin{figure}[t]
\begin{center}
 \includegraphics[width=\textwidth]{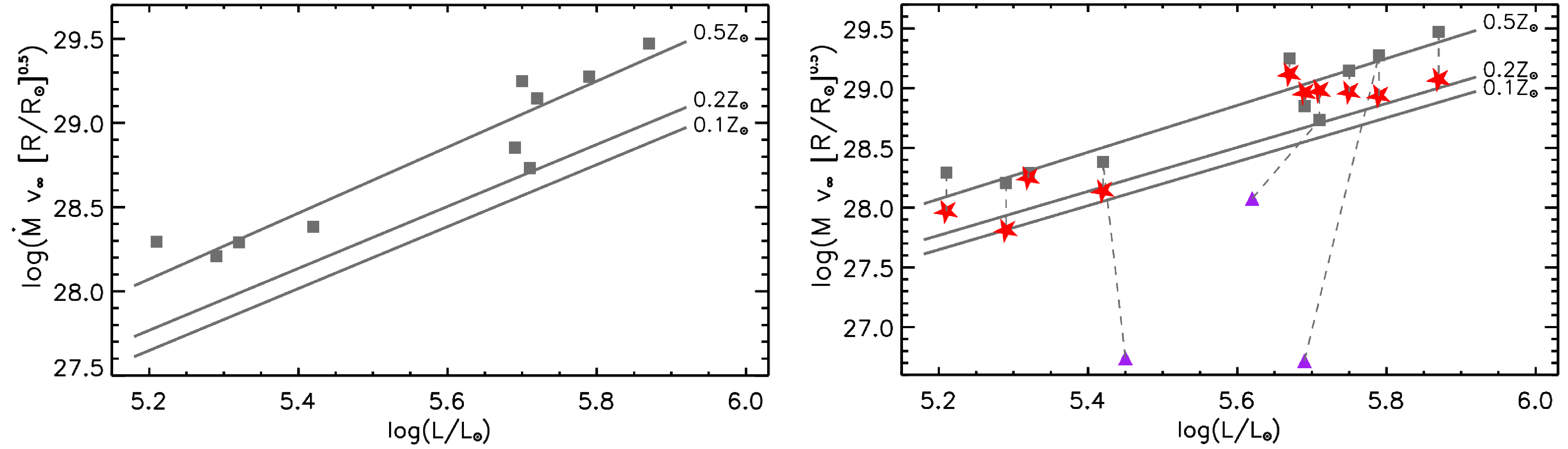} 
 \caption{The 1/7\Osun~ Wind Momentum - Luminosity Relation. 
   \textbf{Left:} Wind-momentum of 1/7\Osun~ OB-stars from optical data
   \citep{Hal12,Tral11,Tal14}:
   \Mdot's were derived from \halpha, and \vinf=2.65\vesc~ (gray squares).
   The unclumpled observational relations for the MW, LMC and SMC  \citep{Mal07b}
   are shown for comparison.
   \textbf{Right:} The WLR recomputed with
   UV terminal velocities (red stars), either direct measurements of the star or
   a star with similar spectral type \citep[from][]{Gal14,Bal15}.
   The results from a full UV+optical analysis of 3 stars by \citet{Bal15} are also included
   (purple triangles).
 }
   \label{F:WLR}
\end{center}
\end{figure}

When the WLR for 1/7\Osun~ stars was first plotted, the community was alarmed:
most of the stars were found at the loci of more metal-rich 0.5\Zsun~ LMC stars (Fig.\,\ref{F:WLR}-left).
Together with the previous detection of strong P~Cygni profiles in IC~1613's V39 \citep{Hal10},
and the existence of a WO in the galaxy \citep[e.g.][]{Tral15},
these results were suggestive of stronger winds than expected in the studied galaxies.
If these results were confirmed, we would need updated wind recipes, evolutionary
models, feedback calculations and SN and GRB rates for very low-metallicity massive stars.

However, the \textit{strong wind problem} can be largely explained by the lack of suitable
diagnostics for both the wind and metallicity (see Sect.~\ref{ss:meta}).
The optical-only studies that reported the \textit{strong winds}
are based on \halpha~ profile fitting.
They are not sensitive to small mass loss rates and
are degenerated to the parameter 
Q=\Mdot/(\vinf$\rm \cdot R_{\star})^{1.5}$ \citep{KP00}.
In order to derive \Mdot, and lacking a direct ultraviolet (UV) measurement,
\vinf~ is calculated from the escape velocity (\vinf/\vesc=2.65)
and scaled with metallicity \citep[\vinf$\propto$Z$^{0.13}$][]{LRD92}.
The \vinf/\vesc=2.65 calibration is long known to suffer a large scatter \citep{KP00},
but recent results also argue against a straightforward \vinf $\propto$ Z relation \citep{Gal14}.
Thus, in absence of UV data, \vinf~ uncertainties add up and propagate to \Mdot~ and the WLR.

The installation of the Cosmic Origins Spectrograph (COS)
on the HST enabled UV spectroscopy ($\sim$1150-1800\AA)
of OB stars out to $\sim$1Mpc (see Sect.\,\ref{s:sexta}),
providing crucial information on their RDWs.
The wind terminal velocity can be measured from
the resonance lines of
NV~$\lambda$1238.8,1242.8,
SiIV~$\lambda$1393.8,1402.8 and
CIV~$\lambda$1548.2,1550.8.
Together with CIII~$\lambda$1176 and NIV~$\lambda$1718.0,
these lines
constrain mass loss rates even for the low \Mdot$< 10^{-7}$\myr~
values expected for metal-poor stars
that render \halpha~ insensitive.

Fig.\,\ref{F:WLR}-right shows the 1/7\Osun~ WLR
recalculated with UV derived terminal velocities
by \citet{Gal14} and \citet{Bal15}.
The wind-momentum of most of the stars was revised downwards,
and the overall trend is now closer to
\citet{Mal07b}'s regression describing the SMC (0.2\Zsun).
Moreover, the joint UV+optical analysis of 3 stars by \citet{Bal15}
yielded wind-momenta much below the 1/7\Zsun~ line.
While the difference is partly due to the consideration
of clumping in their study, the authors also propose
the existence of a hot gas component in the wind that do not
impact P~Cygni profiles and therefore cannot be diagnosed from the UV.
In any case, the WLR of the 1/7\Osun~ galaxies
is still not settled and clearly requires a larger sample
of high quality spectra.

\subsection{The actual metallicity of IC~1613, NGC~3109 and WLM}
\label{ss:meta}

The HST-COS observations delivered an interesting by-product:
the iron content of IC~1613 and WLM is similar or even larger than the $\sim$1/5\Fesun~ content of the SMC,
superseding the 1/7\Zsun~ value scaled from oxygen
\citep{Gal14,Bal15}.
\citet{Tal07} obtained a similar value from the analysis of three RSGs in IC~1613.
Meanwhile \citet{Hal14} reported an $\sim$0.21\Fesun~ iron abundance  
for the massive stars of NGC~3109.
The updated $\gtrsim$0.2\Fesun~ content of IC~1613, NGC~3109 and WLM
also helps to alleviate the problem of \textit{strong winds},
as iron is the main driver of mass loss \citep[e.g.][]{VKL01} and a larger wind-momentum is expected.

These results reflect the known fact that the ratio of $\alpha$-elements to Fe
depends on the chemical evolution of galaxies,
and that oxygen is not always a good proxy for metallicity.
Unfortunately for the subject of this work,
it also implies that
\textit{the regime of massive
stars with poorer metal-content than the SMC  ($\lesssim$0.1\Fesun) remains unexplored.}

\begin{figure}[]
\begin{center}
 \includegraphics[width=\textwidth]{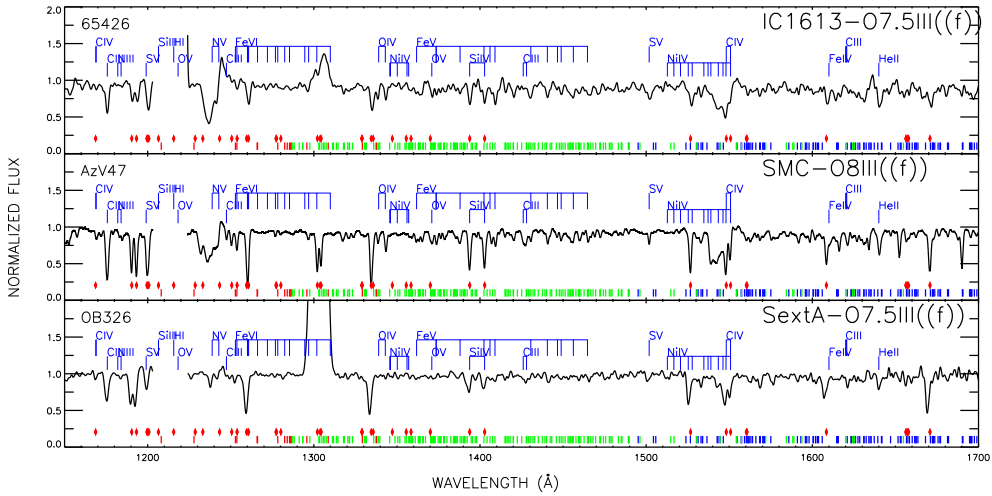} 
 \caption{First UV spectroscopy at sub-SMC metallicity. The chart compares
   HST-COS-G140L observations of stars in IC~1613 and Sextans~A  (PI M. Garcia)
   with the STIS spectrum of a star with similar spectral type.
   \#65426 in IC~1613 shows the strongest P~Cygni profiles of CIV and NV,
   stronger than the SMC star.
   The wind profiles are much weaker in the Sextans~A star,
   as expected with 1/10\Zsun~ metallicity, although the
   terminal velocity can still be measured from CIV.
   The plot also illustrates striking differences in the Fe-content;
   the lines in the FeV forest (green ticks) are strongest in the IC~1613 star,
   and decrease as we move downwards in the plot, suggesting a
   succession of decreasing metallicity.
 }
   \label{F:UV}
\end{center}
\end{figure}

\section{A promising 1/10\Zsun~ galaxy: Sextans~A}
\label{s:sexta}

Located 1.3Mpc away, Sextans~A is potentially the most iron-poor galaxy of the
Local Group.
\citet{Hal14} compiled iron and $\alpha$-element abundances of
nearby irregular galaxies determined from blue supergiants,
thus probing the present-day metallicity while avoiding the calibration
issues of nebular studies:  
Sextans~A stands out as the most iron-poor with [Fe/H]$\sim <$-1.0 and [$\alpha$/Fe]$\sim$0.

 \citet{Cal16} recently confirmed the first OB-stars in Sextans~A (see Table \ref{T:census})
with low-resolution long-slit spectroscopy performed at the GTC.
The data enabled the first determination of the
stellar parameters but yielded no information on the wind or abundances (as expected).
The identified stars are not very massive, \Mini$\sim$20-40\Msun, and 4-6Myr old.
The youngest ones concentrate on the main galactic-scale over-densities of neutral
hydrogen, similarly to IC~1613 \citep{Gal10}, although no very-early spectral types
were detected.
However, the most massive stars of Sextans~A may be
gas- and dust-enshrouded within the HII shells,
and out of reach to the depth of the observing program.

The UV spectra of some Sextans~A OB-stars were obtained by pushing HST-COS
to its limits (Fig.\,\ref{F:UV}). The stars display wind profiles that are weak, but
strong enough to set constraints on the wind properties (Garcia \etal in prep).
The continuum depletion caused by Fe and Ni transitions is almost negligible,
specially when comparing with IC~1613 or SMC stars (see Fig.\,\ref{F:UV}),
therefore confirming the poor Fe content of the stars.

Sextans~A may become the next sub-SMC metallicity standard 
and important observational effort is being devised to mine
this extremely interesting galaxy.
Stay tuned for GTC-OSIRIS multi-object spectroscopic observations (Camacho \etal in prep).

\section{Summary and outlook: prospects for the E-ELT and LUVOIR}

The analysis of stars in the 1/7\Osun~ galaxies
IC~1613, NGC~3109 and WLM has not revealed significant
differences with SMC stars to date. This could be partly due
to the very similar iron content of the four galaxies.
The IC~1613, NGC~3109 and WLM stars open the window
to study the effects of non-solar abundance patterns in the evolution
and winds of massive stars, although such task would require
much larger samples of stars.

The number of confirmed OBA stars
in the true sub-SMC metallicity Sextans~A is still scarce.
The large distance to the galaxy, 1.3Mpc, makes observations challenging.
In addition, internal extinction at the expected sites for OB stars
may be non-negligible since the youngest populations
often co-exist with ionized and neutral gas, and dust.
Deeper observations are needed.

In fact, progress in the characterization of sub-SMC metallicity massive
stars requires efforts in 2 steps. Firstly
deep, low-resolution (R$\sim$ 700-2000), wide-field spectroscopic observations
are key to pierce through gas and dust, reach sufficiently weak stars,
and enlarge the samples.
In this respect, the high multiplexing capabilities of
the second-generation spectrographs on 8-10m telescopes will be crucial.
We can expect significant contributions with MEGARA at the GTC \citep{MEGARA}
and MUSE at the VLT \citep{MUSE} in the near future.

Secondly stellar properties, wind parameters and abundances will be derived
from quantitative spectroscopic analysis. The required
ultraviolet observations and medium resolution (R$\sim$5000)
optical spectroscopy again meet the problem of reaching
stars as faint as V$\sim$20-21 and extinction, critical to the UV.
It is important to stress that UV observations
(optimally in the 950-1800\AA~ range) are a necessary companion
to optical spectroscopy.
This range hosts both the diagnostics
for the terminal velocity of OB stars
and key lines to constrain shocks in the wind \citep{GB04}
and micro-/macro-clumping \citep{SPFO11}.

We are
pushing current observing facilities to the limit.
Not only larger collecting surfaces are needed, but also source
confusion cripples studies of individual objects in interesting sites
farther than Sextans~A.

Beyond the Local Group, I~Zw18 \citep[18.2Mpc][]{Aal07} is the most metal poor galaxy known to
sustain significant star formation.
With 1/32\Zsun~ \citep{VIP98}, it may become the reference to study star formation in the first galaxies
of the Universe
and it is the ideal site to test  1/50\Zsun~ stellar tracks.
In fact, I~Zw18 exhibits intense HeII4686 emission \citep{Keal15} that could be
caused by \citet{Szal15}'s TWUINs, very hot and massive stars experiencing CHE.
However, I~Zw18 is about 10 times farther than Sextans~A
and resolving its stellar population escapes current facilities.

The field will experience a tremendous boost with the arrival of
two great astronomical facilities in the near/mid-term future.
Multi-object spectrographs at the E-ELT (e.g. MOSAIC) will 
be able to reach main-sequence O-stars almost as far as 4Mpc, 
and resolve intricate environments like 30~Dor out to 1.5Mpc.
The second key facility is NASA's likely next flagship mission LUVOIR,
a 12m space telescope with UV, optical and IR capabilities.
The observatory would be 
sensitive to unprecedentedly faint UV sources \textit{per se},
but it will also host a multi-object spectrograph for UV observations (LUMOS)
that will enable to observe entire extragalactic OB-associations in one shot.
Both large facilities will be able to target individual, luminous stars in I~Zw18.
The synergies between E-ELT and LUVOIR will surely  revolutionize
the studies of truly metal-poor massive stars.

\acknowledgments

M. Garcia acknowledges funding by the Spanish MINECO
\textit{via} grants FIS2012-39162-C06-01, ESP2013-47809-C3-1-R and ESP2015-65597-C4-1-R.

\begin{discussion}



\end{discussion}

\end{document}